\documentstyle[preprint,aps]{revtex}
\begin{document}
\draft
\title{ Spectral Functions of Local Operators
In The Tomonaga-Luttinger Model}
\author{S. P. Strong}

\address{
Theoretical Physics, University of Oxford,
1 Keble Rd., Oxford, OX1 3NP, UK \\
and \\
Joseph Henry Laboratories, Princeton University,
P.O. Box 708, Princeton, NJ 08544-0708\\
}
\date{May 20, 1994}
\maketitle
\begin{abstract}
I give a simple  general prescription for computing the
spectral functions of local operators in the Tomonaga-Luttinger
model  from the space-time correlation functions.
The method is significantly simpler than
directly transforming the space-time  Greens function
and allows a physical interpretation of the singularities
encountered in the spectral function.
\end{abstract}

\narrowtext

The Tomonaga-Luttinger model
\cite{Tomonaga,Lutt}
 provides an fascinating
example of a soluble model of an interacting
Fermion problem
with a non-Fermi liquid groundstate.
The low energy eigenstates of the system consist
of bosonic degrees of freedom representing collective
modes of the Fermi surface
\cite{Lieb,LuthPesch,Haldane}.
There are no single particle, Fermionic,
low energy eigenexcitations.
Correspondingly the spectral functions of
local operators in the model display unusual
features including multiple singularities with
non-trivial power laws.  The properties of the
single-electron spectral function
were studied by Meden and Sch\"{o}nhammer
\cite{Meden} and  Voit \cite{Voit}
by Fourier transforming the single-electron
Greens function.
This procedure is entirely correct; however,
it is somewhat unintuitive and the results
take on  complicated form for the case of models
with spin.
I give here a prescription for calculating the
spectral functions of  local operators
from the appropriate space-time Greens functions
based on the observation that many of
the complications
encountered in the Fourier transforming of
the relevant Greens functions arise from the
fact that a general local operator in the
Tomonaga-Luttinger model involves the creation
of four distinct types of bosons
and these complications are
circumvented when the
independent nature of the different boson types is
considered.

 Using forms suitable for
the calculations of spectral
functions in the low energy, long wavelength
limit,
the single electron creation
($\psi_{R~or~L}^{\dagger}(x)$) and annihilation
($\psi_{R~or~L}(x)$)
 operators, the local singlet
($\psi_{R,\uparrow}(x)\psi_{L,\downarrow}(x))$
 and triplet
($\psi_{R,\uparrow}(x)\psi_{L,\uparrow}(x))$
pairing
operators  and the local charge
($\psi^{\dagger}_{R,\uparrow}(x)
\psi_{L,\uparrow}(x)
+
\psi^{\dagger}_{R,\downarrow}(x)
\psi_{L,\downarrow}(x)$)
and spin
($\psi^{\dagger}_{R,\uparrow}(x)\psi_{L,\downarrow}(x) +
\psi^{\dagger}_{L,\uparrow}(x)\psi_{R,\downarrow}(x)$)
 density
wave operators
can all expressed as sums of operators of the form
\cite{LuthPesch,Haldane}:
\begin{equation}
\label{eq:opdef}
\Phi(x;\alpha_{j})
=
\exp(i \sum_j \alpha_j \phi^{\dagger}_j(x))
\exp(i \sum_j \alpha_j \phi_j(x))
\end{equation}
where
\begin{equation}
\label{eq:theta}
\phi_j(x) = (-1)^j \frac{\pi x}{L} N_j + i
\sum_{q \neq 0} \Theta((-1)^j q)
\sqrt{\frac{2 \pi}{L |q|}} e^{-iqx}a_{q,j}
\end{equation}
where $a^{\dagger}_{q,j}$ creates a
density wave excitation of the
$j$ type fermions of the non-interacting model:
\begin{equation}
a^{\dagger}_{q,j} = \sqrt{\frac{2 \pi}{L |q|}}
\sum_k \psi_j^{\dagger}(k+q) \psi_j(k)
\end{equation}
The $j$ index runs from one to four with
with one corresponding to the
 left-moving, up-spin,
two to right-moving, up-spin,
three to left-moving down-spin
and four to right moving down spin
fermions. For the non-interacting
Luttinger model (and in the low energy limit
of the Tomonaga model \cite{Gutfreund}) the
$a^{\dagger}_{q,j}$ create
  eigenexcitations:
\begin{equation}
H_0 = v_F
 \sum_{j,q}(-1)^j \Theta((-1)^j q)   q
a^{\dagger}_{q,j}
a_{q,j}
\end{equation}
In the low energy limit of
the interacting the
Tomonaga-Luttinger model
the eigenexcitations are
linear combinations of the
$a^{\dagger}_{q,j}$
\begin{equation}
\label{eq:ham_int}
H_{int} =  \sum_{j,q}(-1)^j \Theta((-1)^j q)  v_j q
b^{\dagger}_{q,j}
b_{q,j}
\end{equation}
where
\begin{equation}
b_{q,j} = \sum_k
\beta_{jk} (
\frac{1+(-1)^{k-j}}{2} a_{q,k} +
\frac{1-(-1)^{k-j}}{2} a^{\dagger}_{q,k} )
\end{equation}
The original operators can then be expressed
in terms of the new boson operators:
\begin{equation}
\label{eq:opdef2}
\Phi(x;\gamma_{i})
\sim
\exp(i \sum_j \gamma_j \zeta^{\dagger}_j(x))
\exp(i \sum_j \gamma_j \zeta_j(x))
\end{equation}
where
\begin{equation}
\zeta_j(x) =
 \sum_k
\beta_{jk} (
\frac{1+(-1)^{k-j}}{2} \phi_{q,k}(x) +
\frac{1-(-1)^{k-j}}{2} \phi{\dagger}_{q,k}(x) )
\end{equation}
so that
\begin{equation}
\gamma_j = \sum_m \beta_{jm}^{-1} \alpha_m
\end{equation}
where $\beta^{-1}$ is the inverse of $\beta$.

The operator, $\Phi$, is a
product of operators coupling to the
different species of bosons, resulting of
in a product form for the asymptotic,
space-time Greens function \cite{LuthPesch,Haldane,Larkin,Emery}:
\begin{equation}
\label{eq:G}
G(x,t) \sim
(x + v_c t - i \frac{sgn(t)}{\Lambda})^{p_1}
(x - v_c t + i \frac{sgn(t)}{\Lambda})^{p_2}
(x + v_s t - i \frac{sgn(t)}{\Lambda})^{p_3}
(x - v_s t + i \frac{sgn(t)}{\Lambda})^{p_4}
\end{equation}
There is some ambiguity in the choice of
$\beta$'s required to diagonalize the Hamiltonian.
The
choice:
$\beta_{1~or~2,i}=\beta_{1~or~2,i\pm2}$ and
$\beta_{3~or~4,i}=-\beta_{1~or~2,i\pm2}$
is always possible and results in
$p_i = \gamma_i^2$
and, for that case,
$c$ and $s$ subscripts refer to charge (symmetric
combination
of up and down spin bosons)
and spin
(anti-symmetric
combination
of up and down spin bosons)
bosons,
respectively.

The analytic properties of $G$ are quite
complicated and Fourier transforming
directly to find the spectral
function is
generically rather involved even numerically
\cite{Meden} \cite{Voit}.
However, on physical grounds one expects a
significant simplification can be achieved if
on recognizes that the form of $G$
is a simple product resulting from the
the fact that $\Phi$ was a product of
independent operators.
The spectral function of the product of operators
of the form
$\exp(i \sum_j \gamma_j \zeta^{\dagger}_j(x))$
should therefore be given by a convolution of
the spectral functions of the individual
operators:
\begin{eqnarray}
\rho_{\Phi}(k,\omega) & = &
Re~ \frac{1}{\pi}
\int_0^{\infty}dt  e^{i\omega t} ~\int dx e^{-ikx}
\langle GS |\Phi^{\dagger}(x,t)\Phi(0,0)| GS \rangle
 \\ & = &
\sum_m |\langle m |\Phi(0,0)|GS\rangle|^2 \delta(\omega-E_m)
\delta(k-p_m)
\\ & = &
\sum_{m_1,m_2,m_3,m_4}
\left( \prod_j |\langle m_j |
\exp(i \sum_j \gamma_j \zeta^{\dagger}_j(x))
|GS\rangle|^2 \right)
\delta(\omega-\sum_j E_{m_j})
\delta(k-\sum p_{m_j})
 \\ & = &
\int \prod_j d\omega_j
dk_j ~\rho_j(k_j,\omega_j) ~~\delta(k - \sum k_j)
\delta(\omega - \sum \omega_j)
\end{eqnarray}
where
\begin{eqnarray}
\rho_j(k_j,\omega_j) & = &
\sum_{m_j}
|\langle m_j |
\exp(i \sum_j \gamma_j \zeta^{\dagger}_j(x))
|GS\rangle|^2
\delta(\omega_j-E_{m_j})
\delta(k_j-p_{m_j})
\\ & = &
Re~ \frac{1}{\pi}
\int_0^{\infty}dt  e^{i\omega t} ~\int dx e^{-ikx}
\langle GS |
\exp(-i \sum_j \gamma_j \zeta_j((x,t)
\exp(i \sum_j \gamma_j \zeta^{\dagger}_j((0,0)
| GS \rangle
\end{eqnarray}

The utility of this decomposition for the
Tomonaga-Luttinger model results from the
simple form of the spectral functions of the
various parts of $\Phi$. The spectral function
for the operator
$\exp(i \sum_j \gamma_j \zeta^{\dagger}_j(x))$
is given by:
\begin{equation}
\rho_j(k,\omega) =
 |k|^{-1-p_j}
 \delta(\omega - (-1)^j v_j k) \Theta(\omega) e^{-|k|/\Lambda}
\frac{\sin(-p_j \pi )}{\pi} \Gamma(1+p_j)
C(\Lambda)
\end{equation}
 where $C(\Lambda)$ is a cutoff dependent constant
and $v_j$ is the velocity of the type $j$ bosons.

The convolutions are thus significantly simplified by the
presences of multiple delta functions. For $N$ non-zero $\gamma$'s
there are $N$ delta function of this type plus
two overall delta functions for $2N$ variables
of integration. Thus in the worst case
the spectral function involves two
non-trivial integrations, but even these are
only of simple power laws.

For example, for the spinless Tomonaga-Luttinger
model the spectral function for the electron
creation operator involves only one type of
right and
left moving boson and there are no non-trivial
integrations to perform.  The piece of the
spectral function coming from the right Fermi point
is given by:
\begin{eqnarray}
\rho(k,\omega) & \propto & \int_{0}^{\infty} d\omega_1 d\omega_2
 dk_1 dk_2
\delta(\omega-\omega_1-\omega_2) \delta(k + k_1 - k_2)
\delta(\omega_1 - v  k_1) \delta(\omega_2 - v  k_2)
k_1^{-1-p_1} k_2^{-1-p_2} \\
& \propto & \Theta(\omega- v k) \Theta(\omega + v k)
(\omega- v k)^{-1-p_1}
 (\omega + v k)^{-1-p_2}
\end{eqnarray}
where $k$ is measured from $k_F$.
The singularity at the onset frequency arises because at that
frequency it is possible to insert very little momentum into
the left moving bosons and the matrix element for this
is singular as $k \rightarrow 0$.

 This is the general
source of all power law singularities
which arise  in the spectral functions of
local operators in the Tomonaga-Luttinger model: the
divergence of matrix elements to create states with very
little momentum carried by one or more types of bosons.
For instance,
including spin with spin-dependent interactions
so that the electron creation operator coupled to four different
kinds of bosons would result in a singularity at
$\omega = v_c k$ with exponent $-1 - p_1 -p_3 -p_4$,
 resulting from the convolution of
the singularities contributed by the three species of
bosons for which $\omega \neq v_c k$.
(Here the $p$'s are those entering the real space single
electron Green's function and may be obtained from the
Bogoliubov transformation which diagonalizes $H_{int}$
once all interactions have been written in bosonized form.
\cite{fn:psums})
The piece of the creation operator
 spectral function
coming from the right Fermi point
is given by:
\begin{eqnarray}
\rho(k,\omega) & \propto & \int_{0}^{\infty} d\omega_1 d\omega_2
d\omega_3 d\omega_4
dk_1  dk_2 dk_3  dk_4  ~\delta(\omega_1 - v_c  k_1)
\delta(\omega_2 - v_c  k_2)
\delta(\omega_3 - v_s  k_3) \delta(\omega_4 - v_s  k_4)
\\ & &  \nonumber
\delta(\omega-\omega_1-\omega_2 -\omega_3 - \omega_4)
 \delta(k + k_1 - k_2+k_3 -k_4)
k_1^{-1-p_1} k_2^{-1-p_2} k_3^{-1-p_3} k_4^{-1-p_4}
\\ \nonumber
 & & \\
 & \propto &
\label{eq:genrho}
\int_{0}^{\infty} d\omega_1 d\omega_2
\Theta(\omega_1) \Theta(\omega_2 )
\Theta(\omega - v_s k -
(1 + \frac{v_s}{v_c}) \omega_1 -
 (1 - \frac{v_s}{v_c}) \omega_2)
\\ & &  \nonumber
\Theta(\omega + v_s k -
(1 - \frac{v_s}{v_c}) \omega_1 -
 (1 + \frac{v_s}{v_c}) \omega_2)
\\ \nonumber
 & &
\omega_1^{-1-p_1} \omega_2^{-1-p_2}
(\omega - v_s k -
(1 + \frac{v_s}{v_c}) \omega_1 -
 (1 - \frac{v_s}{v_c}) \omega_2)^{-1+p_3}
 \\ & &  \nonumber
(\omega + v_s k -
(1 - \frac{v_s}{v_c}) \omega_1 -
 (1 + \frac{v_s}{v_c}) \omega_2)^{-1+p_4}
\end{eqnarray}
which has exactly the singularity at $\omega = v_c k$
expected.
In general a singularity in the above can occur
for $\omega = v_i k$ and the
exponent of the singularity is always given
by $ -1 - \sum_{j \neq i} p_j$.

 Note that Eq. \ref{eq:genrho}
gives the low energy, long wavelength
spectral function for
any local operator whose real space Green's function
is given by Eq. \ref{eq:G}, so the above formula gives
the spectral function for any local operator in terms of
the exponents and velociites entering into its
space-time Green's function.
The same type of
singularity
at $\omega = v_i k$
is present for all operators with the same
sum of $p_j$'s for $j \neq i$.
It is only
through the
dependence of this and other such sums on the actual
operator that
different operators acquire
different types of singularities.

This method gives a simple, intuitive way of obtaining
the spectral function of all local operators
in the Tomonaga-Luttinger model,
however it suffers from two drawbacks.
The first drawback to this
approach to obtaining the spectral
function arises from the
fact that the inclusion of the cutoff in
Eq.~\ref{eq:G} is not strictly correct.
The
spectral functions obtained from
$G$ will therefore not correctly reproduce cut-off
related features and are, as a result,
in general,
not properly normalized.
Once properly normalized
they should properly reproduce all
of the low energy physics.

The second drawback is that the real part of $G(k,\omega)$ is not
obtained directly in this approach, but requires a further Hilbert
transform of the spectral function.  Since all of the physical
information is encoded in $\rho(k, \omega)$, this is not that
serious a problem, but for numerical applications
where the real part of $G$ is required
it will be a significant
disadvantage for this approach compared with methods
 based on direct Fourier
transformation  of the real space Green's function.

S.~P.~S. acknowledges helpful discussions with
D.~G.~Clarke and P.~W.~Anderson as well
as financial support from
NSF grant DMR-9104873.

\samepage

%
%

%
%

\end{document}